\RequirePackage{fix-cm}
\documentclass[smallextended]{svjour3}       
\smartqed  
\usepackage{graphicx}
\usepackage{multirow}
\usepackage{epsfig}
\journalname{}

\title{Proofs of the Kochen-Specker theorem based on the N-qubit Pauli group}

\author{Mordecai Waegell and P.K. Aravind }

\authorrunning{M.Waegell, P.K. Aravind}

\institute{M.Waegell, P.K. Aravind \at
Physics Department, Worcester Polytechnic Institute, Worcester, MA 01609, U.S.A.\\
\email{caiw@wpi.edu, paravind@wpi.edu}}

\date{\today}

\begin{document}
\maketitle
\begin{abstract}
We present a number of observables-based proofs of the Kochen-Specker (KS) theorem based on the $N$-qubit Pauli group for $N\geq4$, thus adding to the proofs that have been presented earlier for the $2$- and $3$-qubit groups. These proofs have the attractive feature that they can be presented in the form of diagrams from which they are obvious by inspection. They are also irreducible in the sense that they cannot be reduced to smaller proofs by ignoring some subset of qubits and/or observables in them. A simple algorithm is given for transforming any observables-based KS proof into a large number of projectors-based KS proofs; if the observables-based proof has $O$ observables, with each observable occurring in exactly two commuting sets and any two commuting sets having at most one observable in common, the number of associated projectors-based parity proofs is $2^{O}$. We introduce symbols for the observables- and projectors-based KS proofs that capture their important features and also convey a feeling for the enormous variety of both these types of proofs within the $N$-qubit Pauli group. We discuss an infinite family of observables-based proofs whose members include all numbers of qubits from two up, and show how it can be used to generate projectors-based KS proofs involving only nine bases (or experimental contexts) in any dimension of the form $2^{N}$ for $N\geq2$. Some implications of our results are discussed.
\end{abstract}

\section{\label{sec:Intro}Introduction}
In a recent paper \cite{Waegell2012b} we pointed out that the $N$-qubit Pauli group (for $N \geq 2$) is a rich source of both observables-based and projectors-based ``parity proofs" of the Kochen-Specker (KS) theorem \cite{KS1967}. We refer to the proofs as parity proofs because, in either the observables-based or projectors-based versions, they exploit the concept of parity to achieve their ends. The purpose of this paper is to give examples of both types of proofs for $4$- and higher qubit systems and to point out several of their properties that we did not discuss earlier in our work on 2- and 3-qubit systems \cite{Waegell2011b,Waegell2012a}. More precisely, the goals of this paper are the following:\\

\noindent
(1) We explain what we mean by an observables-based KS proof and show how it can be depicted in the form of a diagram from which it is obvious by inspection. The two best known examples of such proofs are a 2-qubit proof due to Peres \cite{Peres1991} and Mermin \cite{Mermin1993} and a 3-qubit proof due to Mermin \cite{Mermin1993} based on earlier work by Greenberger, Horne and Zeilinger \cite{GHZ}. In Ref. \cite{Waegell2012a} we presented several examples of $2$- and $3$-qubit proofs of this kind and in Ref. \cite{Waegell2012b} we indicated that we had found a large number of $4$- and higher qubit proofs but gave few details. Here we give examples of 4-,5- and 6-qubit proofs that convey a feeling for the wide variety of possibilities that open up as one goes to a larger number of qubits. We should stress that we consider only {\it critical} proofs, i.e., ones that cannot be reduced to smaller proofs by omitting some subset of qubits and/or observables in them, but that despite this restriction the number of possibilities still grows very rapidly as one goes to a larger number of qubits.\\

\noindent
(2) We show how any observables-based KS proof can be used to construct a system of projectors and bases from which a large number of projectors-based parity proofs of the KS theorem can be obtained. The simplest example of this procedure is provided by the 2-qubit Peres-Mermin square, whose nine observables give rise to a system of projectors and bases that yield a total of $2^{9} = 512$ projectors-based parity proofs \cite{Waegell2011a,Cabello1996,Kernaghan1994,Pavicic2010}. In recent years a number of other examples of projectors-based parity proofs have been found in four \cite{Waegell2011b,Waegell2011c} and eight \cite{Waegell2012a,KP1995} dimensions. A major point of this paper is that every observables-based KS proof, based on a subset of observables of the Pauli group, gives rise to a system of projectors-based parity proofs, and we give a simple algorithm for making this transition. We illustrate this algorithm in the particular case of a 4-qubit observables-based proof and show how the $2^{12}$ associated projectors-based proofs can be obtained with practically no effort (once the system of projectors and bases within which they are embedded has been set up). In addition to the fact that they are easy to generate, the projectors-based proofs are also easy to check, since only simple counting is called for. Like the observables-based proofs from which they are derived, the projectors-based proofs are critical in the sense that they cannot be whittled down to smaller proofs by omitting some subset of their bases. Because each observables-based proof gives rise to a large number of projectors-based proofs, the variety and quantity of the latter are vastly greater than those of the former.\\

\noindent
(3) It is interesting to ask if there are any infinite classes of observables-based proofs that apply to all numbers of qubits from two up. We have discovered several such classes that we call the Star, the Wheel, the Whorl and the Kite (with the names reflecting the shapes of the associated diagrams). The most complex of these families is the Kite, and we give a detailed discussion of it in this paper. All the members of this family can be represented by diagrams having the form of a kite, with the body consisting of nine observables arranged on a 3 x 3 grid and the tail consisting of a string of observables of arbitrary length. For any number of qubits, a suitable choice of observables (and often more than one) can be placed on the framework of the Kite to yield a KS proof. The $2$-qubit Peres-Mermin square can be regarded as a Kite without a tail, and the higher qubit proofs of this family involve tails of increasing lengths. An interesting feature of this family is that all its members give rise to projectors-based KS proofs involving just nine bases (or experimental contexts), and so are the most compact proofs of this type known in $2^{N}$ dimensions for all $N\geq2$. \\

The next three sections are devoted to a discussion of the above three points. We then comment on the significance of our results and their relation to other work. Although this paper builds upon our earlier work \cite{Waegell2011b,Waegell2012a,Waegell2012b}, it is written to be self contained and requires no familiarity with that work.

\section{\label{sec:2} Observables-based KS proofs}

An observables-based proof of the KS theorem for a system of $N$ qubits consists of a subset of observables of the $N$-qubit Pauli group that forms a number of commuting sets of a special kind. The proof is conveniently displayed in the form of a diagram in which the observables are represented as points (or actually as letters within circles centered at points) and the special commuting sets as lines (which could be straight or curved) joining the points. An observable is represented by a sequence of letters, each of which can be one of $X,Y,Z$ or $I$ (these being the Pauli and identity operators of a qubit). For example, $XYIZZ$ represents a 5-qubit observable that is the tensor product of the observables $X,Y,I,Z$ and $Z$ of the individual qubits. Every special commuting set in any of our proofs has the property that the product of all the observables in it is either {+\bf I} or {$-$\bf I}, where {\bf I} is the identity operator in the space of all the qubits. Sets with product {+\bf I} are shown by thin lines and sets with product {$-$\bf I} by thick lines in our diagrams. Any diagram representing an observables-based KS proof has the following two properties: (A) each observable lies at the intersection of an even number of lines, and (B) the total number of thick lines is odd. These properties guarantee that the diagrams provide proofs of the KS theorem. To see why, note that the eigenvalues of any $N$-qubit observable are $\pm 1$ and that a noncontextual hidden variables theory is required to assign the value $+1$ or $-1$ to each of the observables in such a way that the product of the values assigned to the observables on a thin (or a thick) line equals $+1$ (or $-1$). However properties (A) and (B) rule out such a value assignment\footnote{This can be seen as follows. Let $v_{\alpha}$ be the product of the values assigned to the observables in the commuting set indexed by $\alpha$ and consider the product $P = \prod v_{\alpha}$ taken over all the commuting sets. Property (B) requires that $P = -1$ but property (A) requires that $P = +1$, so there is a contradiction.} and so prove the KS theorem.\\

A number of $N$-qubit proof-diagrams are shown in Figs.1-4 for $N = 4,5$ and $6$. The names we have given to the diagrams are whimsical and merely try to capture their shapes. We have also attached a symbol to each diagram that should help the reader pick out the commuting sets in it (particularly in the case of the more complicated diagrams). The left half of each symbol lists the number of observables of each multiplicity (with the multiplicities as subscripts) and the right half lists the number of commuting sets of each size (with the sizes as subscripts). For example, the symbol  ${\bf 12_{2}}$-${\bf1_{5}4_{4}1_{3}}$ for the 4-qubit Star of Fig.1 indicates that it contains 12 observables of multiplicity 2 (i.e., that each occur twice among its commuting sets) and that there are four commuting sets of four observables and one each of five and three observables. The sum of the products of each number with its subscript in the left half of the symbol must equal the similar sum of products on the right, and this can be used as a quick check on the consistency of the symbol. For a symbol to represent a valid KS proof, all the subscripts in its left half must be even (it is also necessary that the number of commuting sets with product {$-$\bf I} be odd, but this fact is not made evident in the symbol and can only be checked by looking at the diagram.) \\

\begin{figure}
\centering
\begin{tabular}{cc}
\epsfig{file=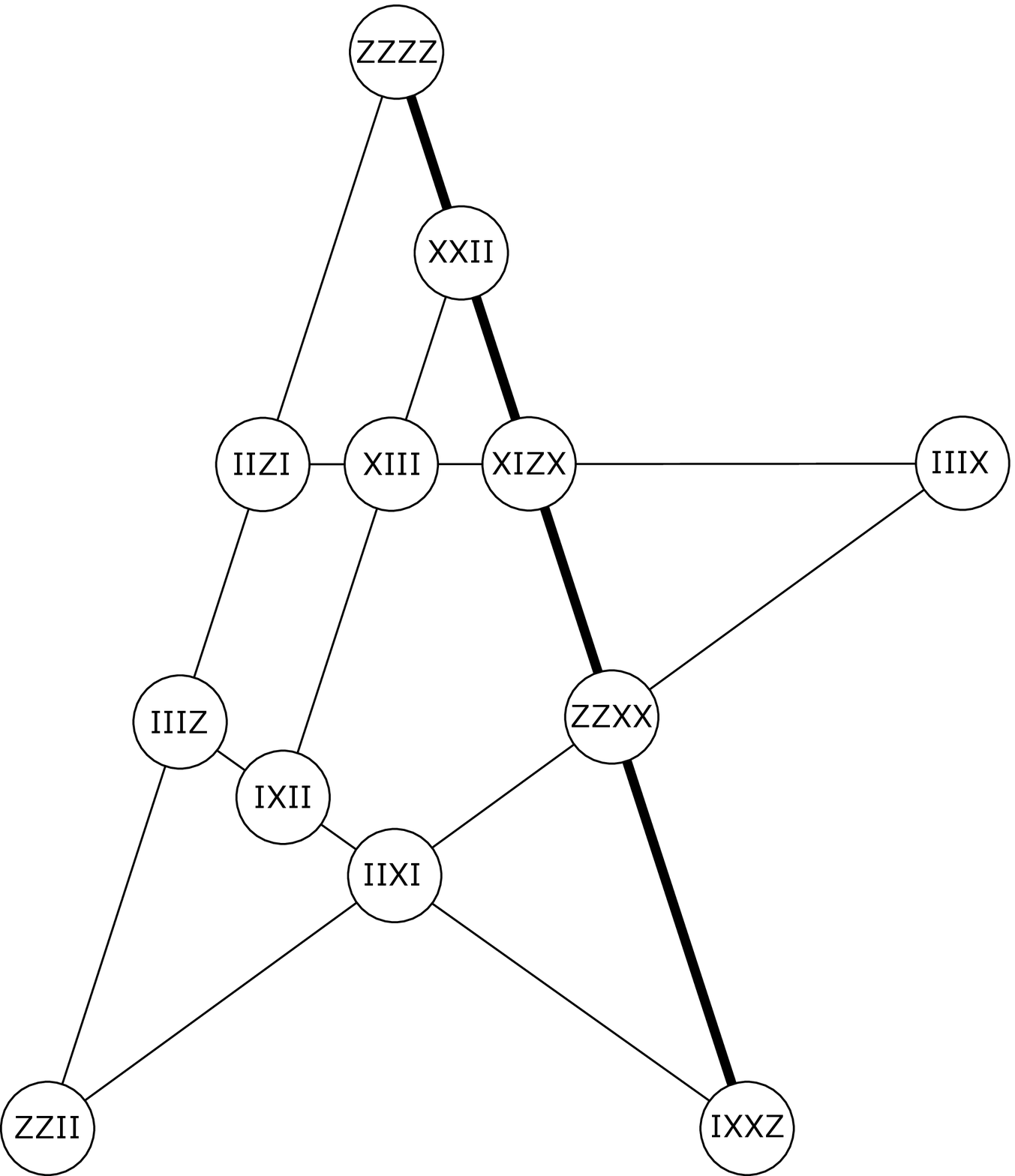,width=0.5\linewidth,clip=} &
\epsfig{file=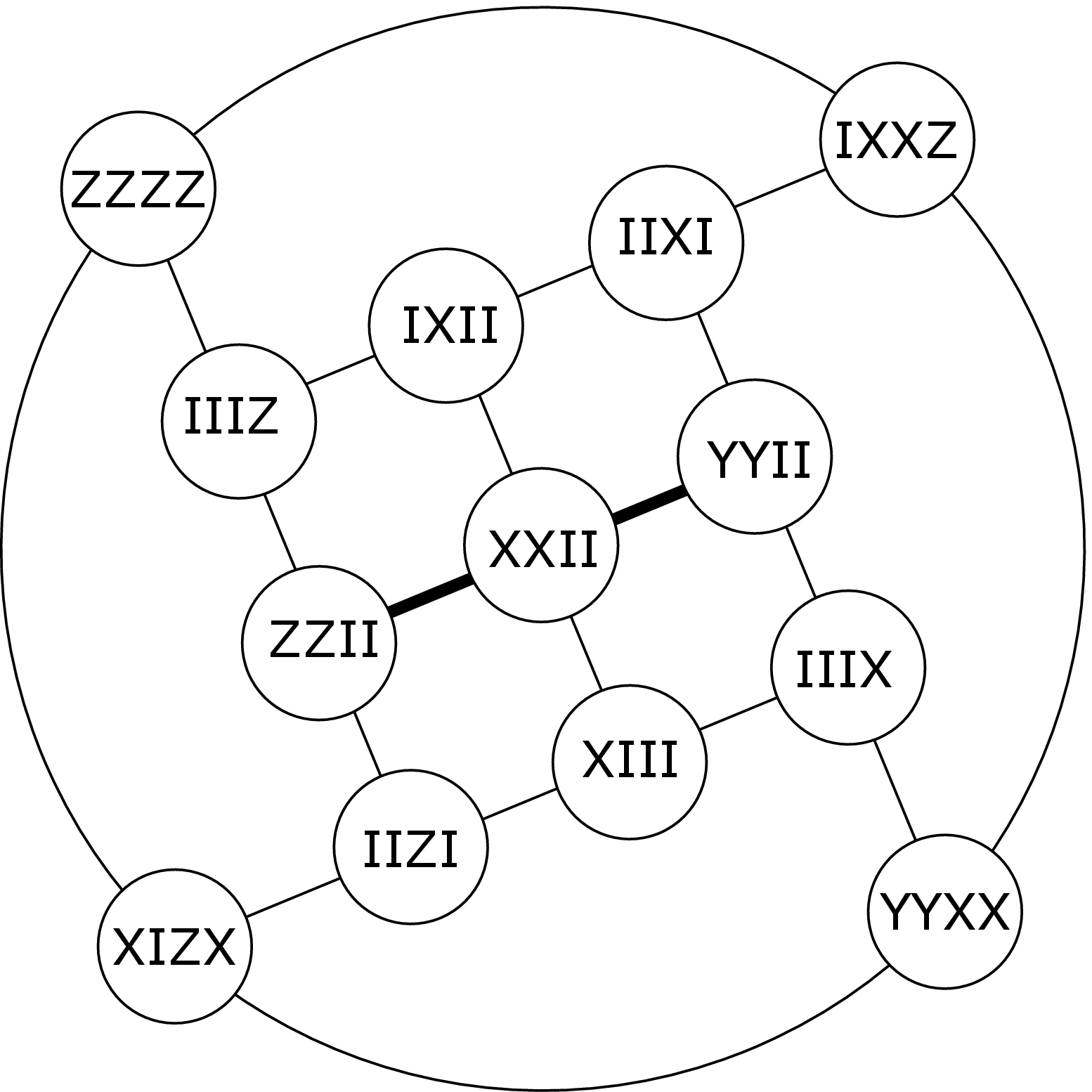,width=0.4\linewidth,clip=} \\
\end{tabular}
\caption{4-qubit Star, ${\bf 12_{2}}$-${\bf1_{5}4_{4}1_{3}}$ (left) and 4-qubit Windmill, ${\bf 13_{2}}$-${\bf 5_{4}2_{3}}$ (right).}
\label{Fig.1}
\end{figure}

\begin{figure}
\centering
\begin{tabular}{cc}
\epsfig{file=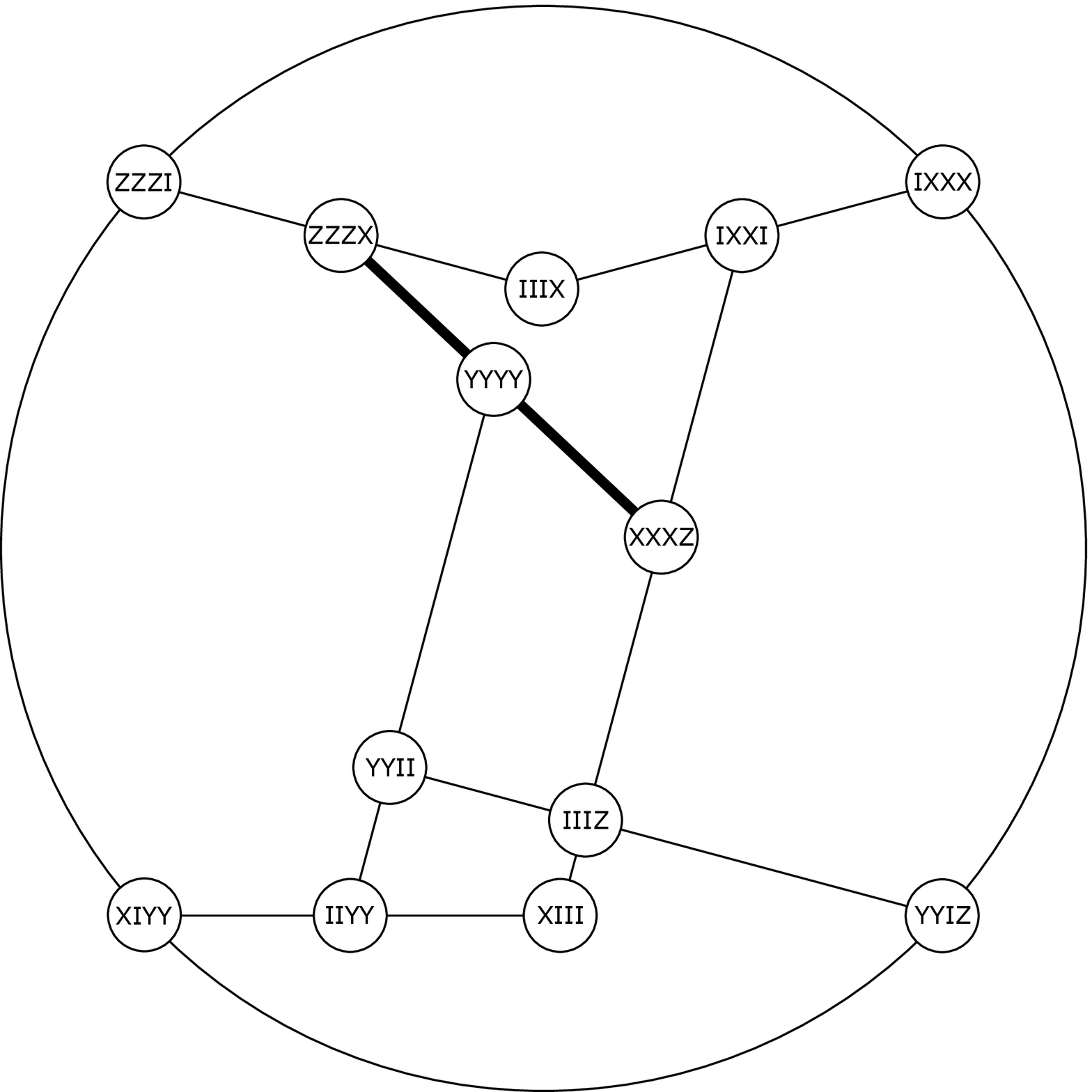,width=0.5\linewidth,clip=} &
\epsfig{file=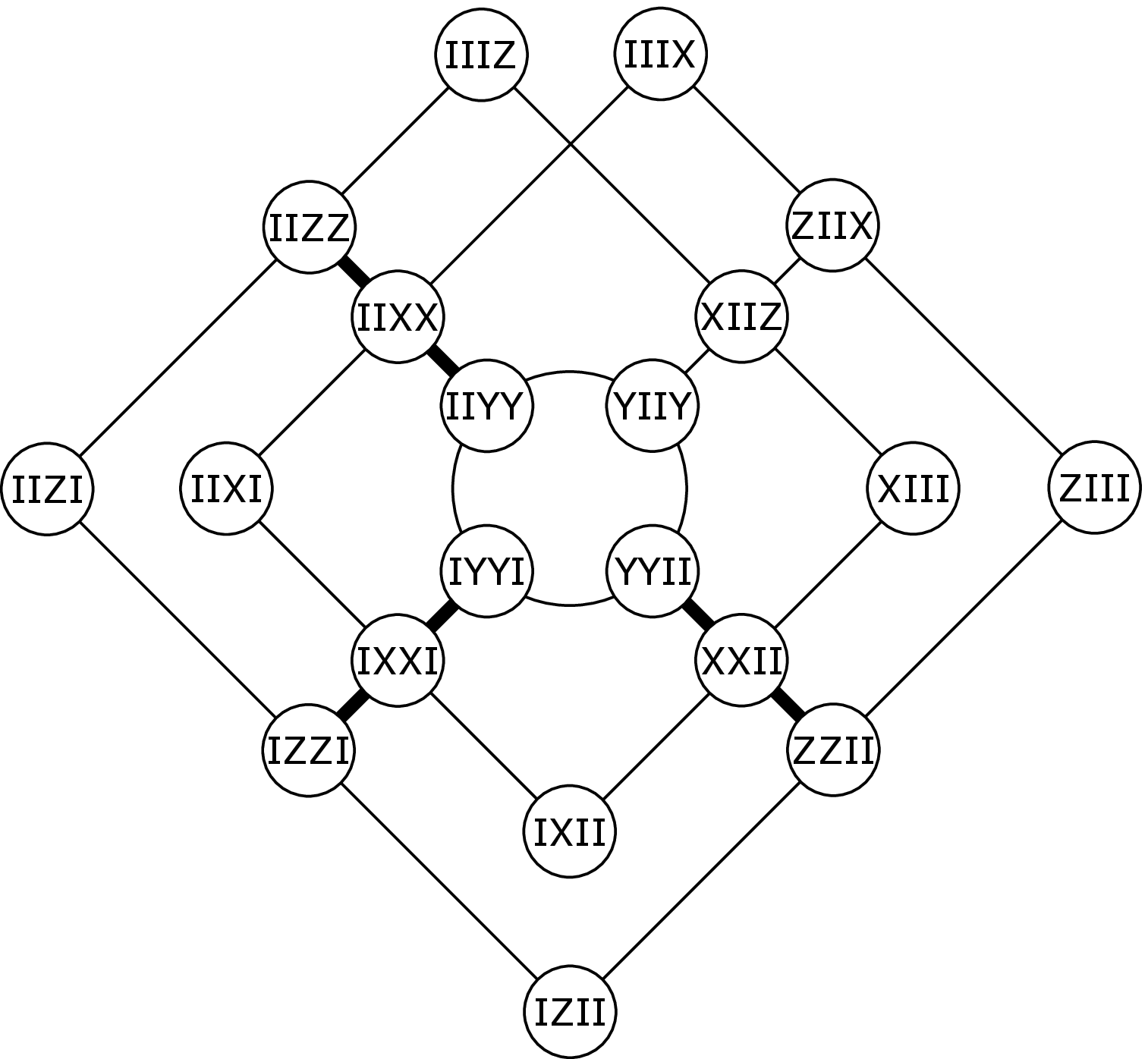,width=0.5\linewidth,clip=} \\
\end{tabular}
\caption{4-qubit Clock, ${\bf 13_{2}}$-${\bf2_{4}6_{3}}$ (left) and 4-qubit Whorl, ${\bf 20_{2}}$-${\bf 1_{4}12_{3}}$ (right).}
\label{Fig.2}
\end{figure}

\begin{figure}
\centering
\begin{tabular}{cc}
\epsfig{file=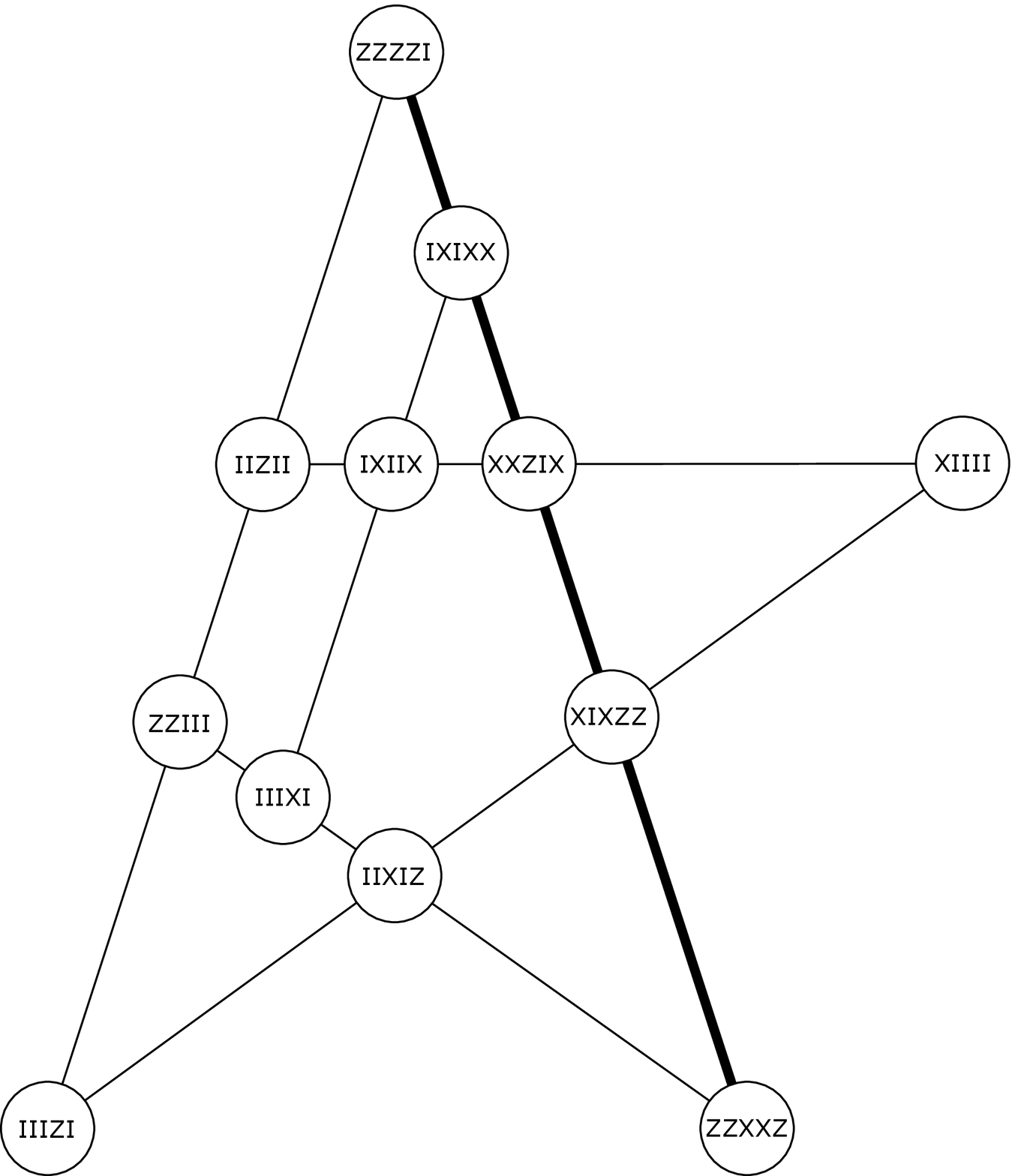,width=0.5\linewidth,clip=} &
\epsfig{file=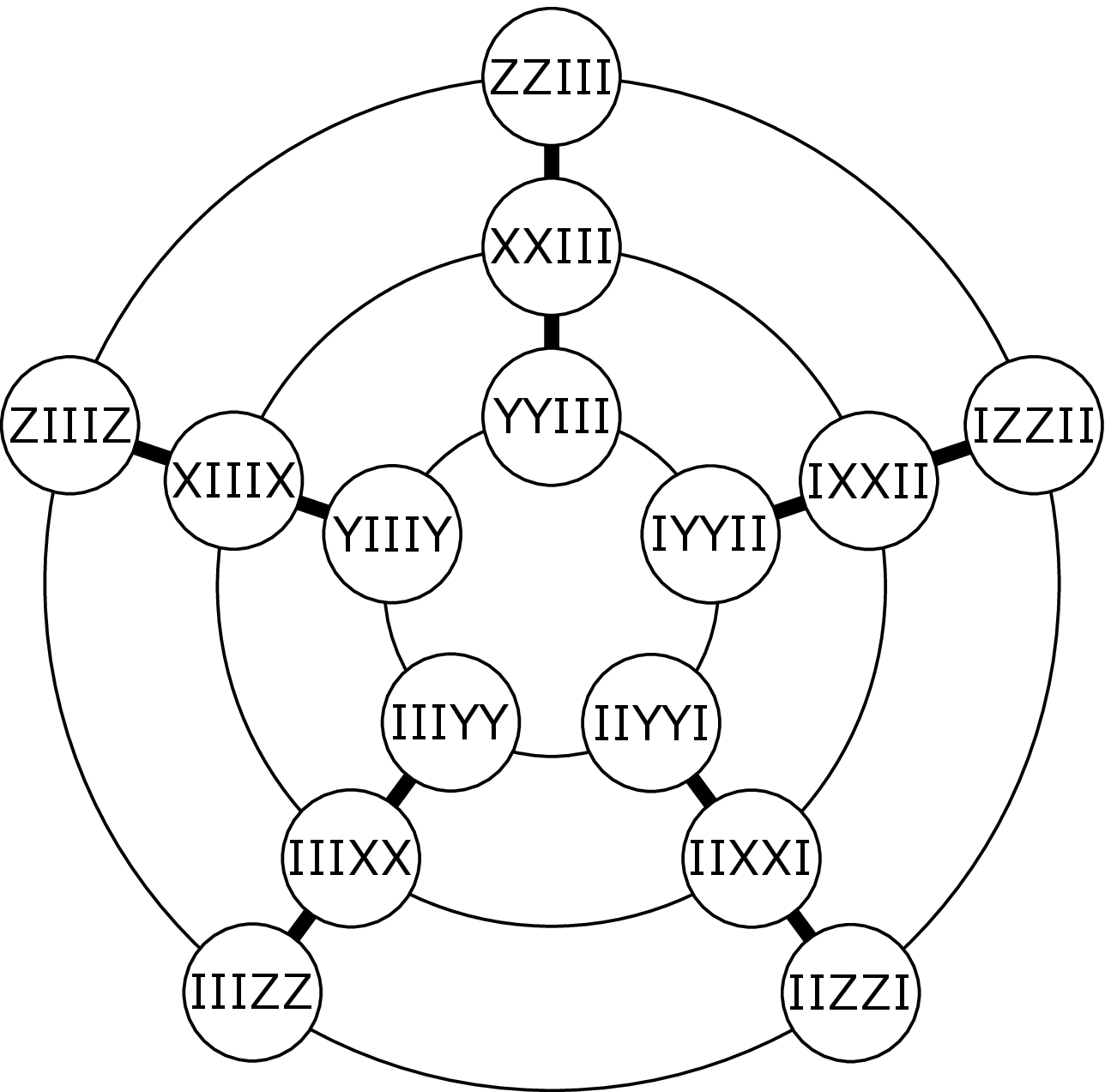,width=0.4\linewidth,clip=} \\
\end{tabular}
\caption{5-qubit Star, ${\bf 12_{2}}$-${\bf1_{5}4_{4}1_{3}}$ (left) and 5-qubit Wheel, ${\bf 15_{2}}$-${\bf 3_{5}5_{3}}$ (right).}
\label{Fig.2}
\end{figure}

\begin{figure}
\centering
\begin{tabular}{cc}
\epsfig{file=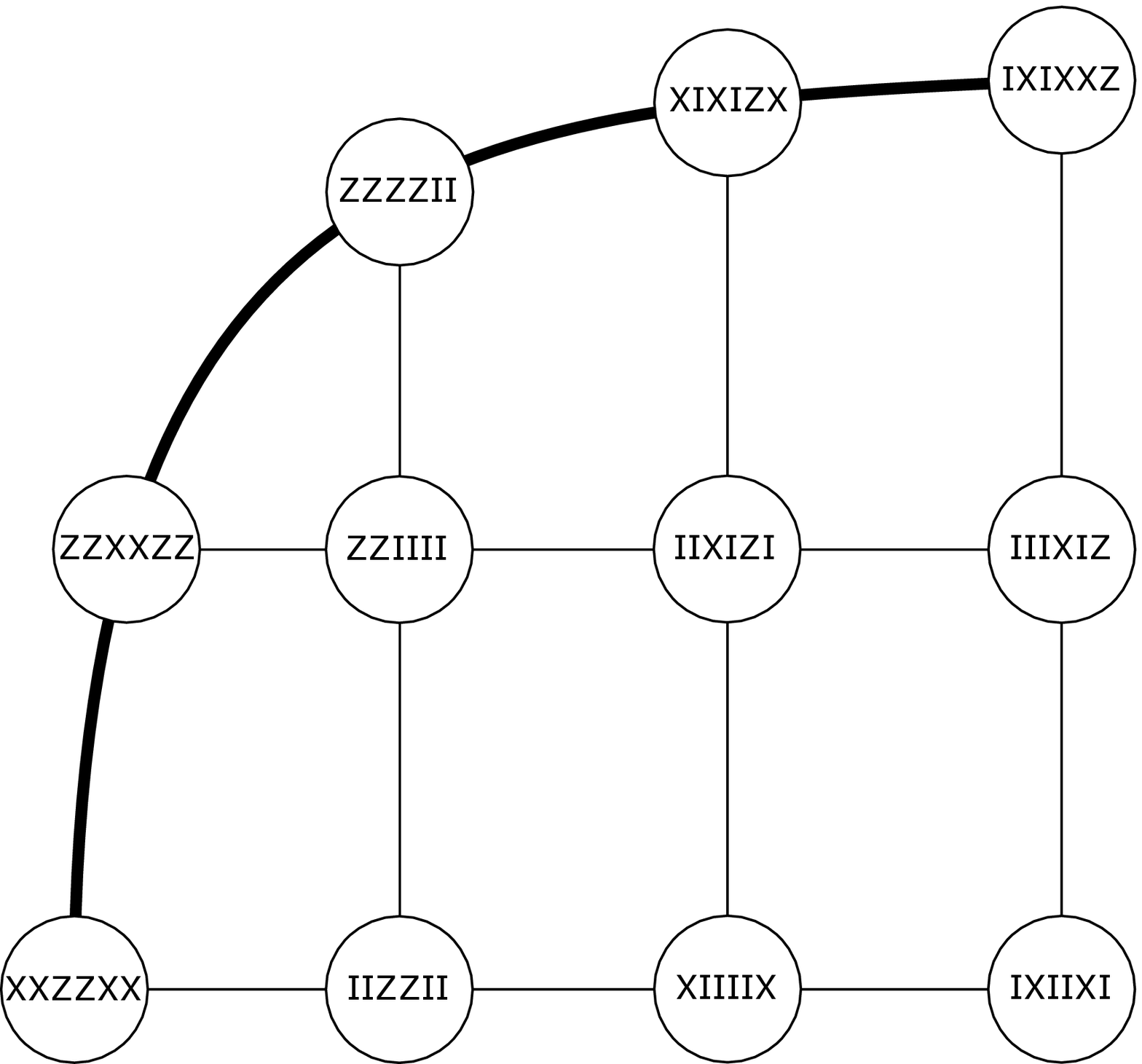,width=0.4\linewidth,clip=} &
\epsfig{file=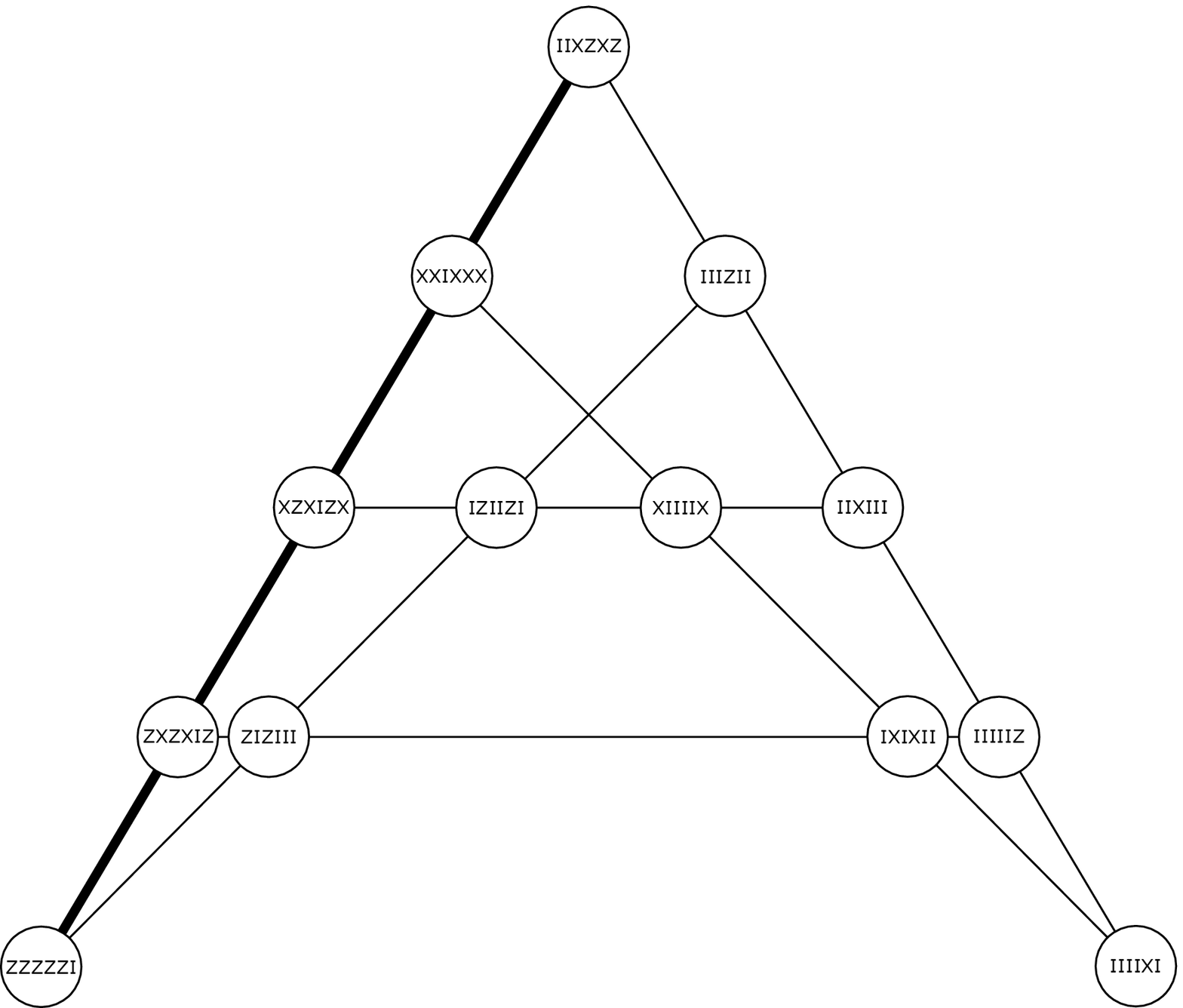,width=0.7\linewidth,clip=} \\
\end{tabular}
\caption{6-qubit Arch, ${\bf 11_{2}}$-${\bf1_{5}2_{4}3_{3}}$ (left) and 6-qubit Arrow, ${\bf 13_{2}}$-${\bf 2_{5}4_{4}}$ (right).}
\label{Fig.2}
\end{figure}

The proof diagrams we have shown here are just a small fraction of the ones we have discovered. We hope to make a more extensive collection available at a website we plan to set up.

\section{\label{sec:3} Projectors-based KS proofs}

Each of the observables-based proofs of the previous section can be used to generate a large number of projectors-based proofs of the KS theorem. We illustrate how this can be done by considering one particular case in detail, namely, the 4-qubit Star of Fig.1. The procedure for obtaining the projectors-based proofs is as follows:\\
\noindent
\\
(1) First enumerate the projectors that are the simultaneous eigenstates of the various sets of commuting observables in the proof. Each commuting set defines a number of mutually orthogonal projectors that sum to the identity and that we term a ``pure" basis. The 4-qubit Star consists of six commuting sets and so gives rise to $6$ pure bases. However these bases are of different sizes, with four consisting of 8 rank-2 projectors, one of 16 rank-1 projectors and one of 4 rank-4 projectors. We now establish a numbering scheme for the projectors. To do this we label the projectors of each commuting set by their eigenvalues with respect to the observables of that set and then convert the eigenvalue strings into binary strings by the replacement $+1 \rightarrow 0, -1 \rightarrow 1$ and finally arrange the binary strings in ascending order after ignoring the digit at the extreme right. With the projectors within each commuting set ordered in this fashion, we then number the projectors sequentially from 1 up, beginning with the first commuting set and proceeding to the others. This procedure is illustrated in Table 1 for the 4-qubit Star, whose commuting sets are shown in the first column. As an example of the numbering procedure, the projectors corresponding to the observables in the last row are represented by the binary strings 000, 011, 101 and 110 (which are arranged in ascending order according to their first two digits) and assigned the numbers 49 to 52, respectively (since the numbers 1 to 48 have already been taken by the earlier projectors). \\

\begin{table}[ht]
\centering 
\begin{tabular}{|c | c | c |} 
\hline 
 Observables & Projectors  &  Rank  \\
\hline
$ZZZZ$, $ZZXX$, $XXII$, $XIZX$, $IXXZ$ & $1-16$ & $1$  \\
$ZZZZ$, $ZZII$, $IIZI$, $IIIZ$          & $17-24$ & $2$   \\
$ZZXX$, $ZZII$, $IIXI$, $IIIX$           & $25-32$ & $2$   \\
$XIZX$, $IIZI$, $XIII$, $IIIX$         & $33-40$ & $2$   \\
$IXXZ$, $IIIZ$, $IXII$, $IIXI$         & $41-48$ & $2$   \\
$XXII$, $XIII$, $IXII$         & $49-52$ & $4$   \\
\hline
\end{tabular}
\caption{The projectors of the 4-qubit Star. The projectors are the simultaneous eigenstates of the mutually commuting observables in each of the rows, and are numbered as explained in the text. The product of the eigenvalue signatures of the projectors in the first row is $-1$, while it is $+1$ for the projectors in each of the last five rows. The ranks of the projectors associated with each commuting set are shown in the last column.}
\label{table1} 
\end{table}

\begin{table}[ht]
\centering 
\begin{tabular}{|c | c c c c c c c c c c c c c c c c |} 
\hline 
Index &\multicolumn{16}{|c|}{Projectors in Basis}\\
\hline
1 & 1 & 2 & 3 & 4 & 5 & 6 & 7 & 8&9&10&11&12&13&14&15&16 \\
2 & 17&18&19&20&21&22&23&24&&&&&&&& \\
3 & 25&26&27&28&29&30&31&32&&&&&&&& \\
4 & 33&34&35&36&37&38&39&40&&&&&&&& \\
5 & 41&42&43&44&45&46&47&48&&&&&&&& \\
6 & 49&50&51&52&&&&&&&&&&&& \\
\hline
7a&1&2&3&4&5&6&7&8&21&22&23&24&&&&\\
7b&9&10&11&12&13&14&15&16&17&18&19&20&&&&\\
8a&1&2&3&4&9&10&11&12&29&30&31&32&&&&\\
8b&5&6&7&8&13&14&15&16&25&26&27&28&&&&\\
9a&1&3&5&7&9&11&13&15&37&38&39&40&&&&\\
9b&2&4&6&8&10&12&14&16&33&34&35&36&&&&\\
10a&1&4&6&7&10&11&13&16&41&42&43&44&&&&\\
10b&2&3&5&8&9&12&14&15&45&46&47&48&&&&\\
11a&1&2&5&6&9&10&13&14&51&52&&&&&&\\
11b&3&4&7&8&11&12&15&16&49&50&&&&&&\\
12a&17&18&21&22&27&28&31&32&&&&&&&&\\
12b&19&20&23&24&25&26&29&30&&&&&&&&\\
13a&17&19&21&23&35&36&39&40&&&&&&&&\\
13b&18&20&22&24&33&34&37&38&&&&&&&&\\
14a&17&20&22&23&43&44&47&48&&&&&&&&\\
14b&18&19&21&24&41&42&45&46&&&&&&&&\\
15a&25&28&30&31&34&35&37&40&&&&&&&&\\
15b&26&27&29&32&33&36&38&39&&&&&&&&\\
16a&25&27&29&31&42&43&45&48&&&&&&&&\\
16b&26&	28&	30&	32&	41&	44&	46&	47&&&&&&&&\\
17a&33&35&37&39&50&52&&&&&&&&&&\\
17b&34&36&38&40&49&51&&&&&&&&&&\\
18a&41&43&45&47&50&51&&&&&&&&&&\\
18b&42&44&46&48&49&52&&&&&&&&&&\\
\hline
\end{tabular}
\caption{The 30 bases formed by the 52 projectors of the 4-qubit Star. They consist of six pure bases (shown above the line) and 24 hybrid bases (shown below the line), with the bases numbered as shown at the left. The hybrid bases come in complementary pairs, with the members of a pair bearing the same number and being distinguished by the letters a and b.}
\label{table2} 
\end{table}
\noindent
(2) In addition to the pure bases, the projectors form a number of ``hybrid" bases that consist of mixtures of projectors from different pure bases (the hybrid bases, like the pure bases, consist of sets of mutually orthogonal projectors that sum to the identity). In order to construct the hybrids, it is necessary to be able to pick out orthogonalities between projectors belonging to different pure bases. This can be done by using the following rule: two projectors from different pure bases are orthogonal if and only if they are  eigenstates of one or more common observables with differing eigenvalues for at least one of those observables. As an example, projector 3 of the first row of Table 1 (which is represented by the binary string 00101) is orthogonal to projector 21 of the second row (which is represented by the string 1001) because they have opposite eigenvalues for the observable $ZZZZ$.  Using this rule it is easy to pick out all the hybrid bases formed by the projectors in Table 1 and these are listed, along with the six pure bases, in Table 2. Note that each hybrid basis is made up of the halves of two pure bases, whose other halves make up a second hybrid complementary to the first one. Complementary hybrids are listed next to each other in Table 2 and bear the same number, but are distinguished by the letters a and b.\\

\begin{table}[ht]
\centering 
\begin{tabular}{|c | c | c | c | c |} 
\hline 
 Index & Projectors  &  Bases & Symbol & Count  \\
\hline
&&&&\\
$1$ & $47$ & $13$ & $5^{1}_{2}10^{1}_{4}1^{1}_{6}24^{2}_{2}4^{2}_{4}3^{4}_{2}-1_{16}4_{12}1_{10}5_{8}2_{6}$ & 128\\
&&&&\\
$2$ & $47$ & $13$ & $10^{1}_{2}5^{1}_{4}22^{2}_{2}7^{2}_{4}3^{4}_{2}-4_{12}1_{10}6_{8}2_{6}$ & 512\\
&&&&\\
$3$ & $47$ & $13$ & $10^{1}_{2}5^{1}_{4}24^{2}_{2}4^{2}_{4}3^{4}_{2}1^{4}_{4}-4_{12}1_{10}5_{8}2_{6}1_{4}$ & 128\\
&&&&\\
$4$ & $49$ & $15$ & $5^{1}_{2}10^{1}_{4}1^{1}_{6}20^{2}_{2}10^{2}_{4}3^{4}_{2}-1_{16}4_{12}1_{10}7_{8}2_{6}$ & 768\\
&&&&\\
$5$ & $49$ & $15$ & $5^{1}_{2}10^{1}_{4}1^{1}_{6}22^{2}_{2}7^{2}_{4}3^{4}_{2}1^{4}_{4}-1_{16}4_{12}1_{10}6_{8}2_{6}1_{4}$ & 512\\
&&&&\\
$6$ & $49$ & $15$ & $10^{1}_{2}5^{1}_{4}18^{2}_{2}13^{2}_{4}3^{4}_{2}-4_{12}1_{10}8_{8}2_{6}$ & 512\\
&&&&\\
$7$ & $49$ & $15$ & $10^{1}_{2}5^{1}_{4}20^{2}_{2}10^{2}_{4}3^{4}_{2}1^{4}_{4}-4_{12}1_{10}7_{8}2_{6}1_{4}$ & 768\\
&&&&\\
$8$ & $51$ & $17$ & $5^{1}_{2}10^{1}_{4}1^{1}_{6}16^{2}_{2}16^{2}_{4}3^{4}_{2}-1_{16}4_{12}1_{10}9_{8}2_{6}$ & 128\\
&&&&\\
$9$ & $51$ & $17$ & $5^{1}_{2}10^{1}_{4}1^{1}_{6}18^{2}_{2}13^{2}_{4}3^{4}_{2}1^{4}_{4}-1_{16}4_{12}1_{10}8_{8}2_{6}1_{4}$ & 512\\
&&&&\\
$10$ & $51$ & $17$ & $10^{1}_{2}5^{1}_{4}16^{2}_{2}16^{2}_{4}3^{4}_{2}1^{4}_{4}-4_{12}1_{10}9_{8}2_{6}1_{4}$ & 128\\
&&&&\\
\hline
\end{tabular}
\caption{The ten different types of projectors-based KS proofs contained in the 4-qubit Star of Fig.1. The second and third columns give the number of projectors and bases in each proof, while the fourth column gives the detailed symbol of the proof (see text for explanation). The fifth column lists the number of distinct proofs of each type, with the sum of all the numbers in this column being $4096 = 2^{12}$.}
\label{table3} 
\end{table}

\begin{table}[ht]
\centering 
\begin{tabular}{|c | c c c c c c c c c c c c c c c c c|} 
\hline 
 Index & \multicolumn{17}{|c|}{Bases in example proof} \\
\hline
$1$ & $7a$ & $8a$ & $9a$ & $10a$ & $11a$ & $12a$ & $13a$ & $14b$ & $15a$ & $16a$ & $17b$ & $18b$ & $1$ & $$ & $$ & $$ & $$ \\
$2$ & $7a$ & $8a$ & $9a$ & $10a$ & $11b$ & $12a$ & $13a$ & $14a$ & $15a$ & $16a$ & $17b$ & $18a$ & $2$ & $$ & $$ & $$ & $$ \\
$3$ & $7a$ & $8a$ & $9a$ & $10a$ & $11b$ & $12a$ & $13a$ & $14b$ & $15a$ & $16a$ & $17b$ & $18b$ & $6$ & $$ & $$ & $$ & $$ \\
$4$ & $7a$ & $8a$ & $9a$ & $10a$ & $11a$ & $12a$ & $13a$ & $14a$ & $15a$ & $16a$ & $17a$ & $18a$ & $1$ & $2$ & $4$ & $$ & $$ \\
$5$ & $7a$ & $8a$ & $9a$ & $10a$ & $11a$ & $12a$ & $13a$ & $14a$ & $15a$ & $16a$ & $17b$ & $18a$ & $1$ & $2$ & $6$ & $$ & $$ \\
$6$ & $7a$ & $8a$ & $9a$ & $10a$ & $11b$ & $12a$ & $13a$ & $14a$ & $15a$ & $16a$ & $17a$ & $18b$ & $2$ & $4$ & $5$ & $$ & $$ \\
$7$ & $7a$ & $8a$ & $9a$ & $10a$ & $11b$ & $12a$ & $13a$ & $14a$ & $15a$ & $16a$ & $17a$ & $18a$ & $2$ & $4$ & $6$ & $$ & $$ \\
$8$ & $7a$ & $8a$ & $9a$ & $10a$ & $11a$ & $12a$ & $13a$ & $14a$ & $15a$ & $16b$ & $17a$ & $18a$ & $1$ & $2$ & $3$ & $4$ & $5$ \\
$9$ & $7a$ & $8a$ & $9a$ & $10a$ & $11a$ & $12a$ & $13a$ & $14a$ & $15a$ & $16a$ & $17a$ & $18b$ & $1$ & $2$ & $4$ & $5$ & $6$ \\
$10$ & $7a$ & $8a$ & $9a$ & $10a$ & $11b$ & $12a$ & $13a$ & $14a$ & $15a$ & $16b$ & $17a$ & $18a$ & $2$ & $3$ & $4$ & $5$ & $6$ \\
\hline
\end{tabular}
\caption{One example of each of the 10 different types of projectors-based proofs listed in Table 3. The bases in each proof are labeled using the notation of Table 2.}
\label{table4} 
\end{table}

\noindent
(3) The system of projectors and bases yielded by any observables-based KS proof contains a large number of projectors-based parity proofs. Any projectors-based parity proof consists of an odd number of bases with the property that each of the projectors that occurs in them occurs in an even number of them. This condition guarantees that it is impossible to assign noncontextual 0/1 values to the projectors in such a way that each basis has exactly one projector assigned the value 1 in it, which proves the KS theorem.\\

\noindent
It is useful to have a symbol for the system of projectors and bases that results from any observables-based KS proof. We use a symbol consisting of two halves, with the left half listing the properties of the projectors and the right half the properties of the bases. Each number in the left half represents the number of projectors of a particular rank and multiplicity (with the rank indicated as a superscript and the multiplicity as a subscript), and each number in the right half represents the number of bases of a particular size (with the size indicated as a subscript). As an example, the symbol for the system in Table 2 is $16^{1}_{6}32^{2}_{5}4^{4}_{4}-1_{16}8_{12}2_{10}14_{8}4_{6}1_{4}$ and it indicates, among other things, that there are 32 rank-2 projectors of multiplicity five and 14 bases of eight projectors in this system. We use a similar symbol to denote any projectors-based parity proof.\footnote{Note that we use bold font for the symbols of the observables-based proofs and ordinary font for the symbols of the projectors-based proofs in order to avoid any confusion between them.} \\

\noindent
We are now in a position to explain how all the projectors-based proofs listed in Table 3 can be picked out from the bases in Table 2. All one has to do is to pick one member from each pair of complementary hybrids (which can be done in $2^{12}$ ways) and supplement them with the needed pure bases to complete the proof. As an example, suppose one picks the 12 hybrids shown in the first row of Table 4. One finds that all the projectors that occur in these bases occur an even number of times among them, with the exception of projectors 1 through 16, which each occur either once or thrice; it is then clear that one should pick pure basis 1 to ensure that all the projectors occur an even number of times among the bases (and also that the total number of bases is odd). This yields the proof shown in the first row of Table 4, whose symbol is indicated in the first row of Table 3 . There are 128 different proofs of this kind, as noted in the last column of Table 3. By picking all possible combinations of hybrid bases, it is possible to generate all the proofs listed in Table 3. It is interesting to note that while there are several proofs involving the same total number of projectors and bases, their detailed structure (as revealed by their symbols) is quite different.\\

\begin{table}[ht]
\centering 
\begin{tabular}{|c | c | c | c | c |} 
\hline 
 Proof & Diagram  &  Symbol & Pure/Hybrid bases & Parity proofs\\
\hline
&&&&\\
4-qubit Star & Fig.1 (left) & ${\bf 12_{2}}$-${\bf1_{5}4_{4}1_{3}}$ & $6/24$ & $2^{12}=4096$\\
&&&&\\
4-qubit Windmill & Fig.1 (right) & ${\bf 13_{2}}$-${\bf 5_{4}2_{3}}$ & $7/26$ & $2^{13}=8192$\\
&&&&\\
4-qubit Clock & Fig.2 (left) & ${\bf 13_{2}}$-${\bf2_{4}6_{3}}$  & $8/26$ & $2^{13}=8192$\\
&&&&\\
4-qubit Whorl & Fig.2 (right) & ${\bf 20_{2}}$-${\bf 1_{4}12_{3}}$ & $13/40$ & $2^{20}=1048576$\\
&&&&\\
5-qubit Star & Fig.3 (left) & ${\bf 12_{2}}$-${\bf1_{5}4_{4}1_{3}}$ & $6/24$ & $2^{12}=4096$\\
&&&&\\
5-qubit Wheel & Fig.3 (right) & ${\bf 15_{2}}$-${\bf 3_{5}5_{3}}$ & $8/30$ & $2^{15}=32768$\\
&&&&\\
6-qubit Arch & Fig.4 (left) & ${\bf 11_{2}}$-${\bf1_{5}2_{4}3_{3}}$ & $6/22$ & $2^{11}=2048$\\
&&&&\\
6-qubit Arrow & Fig.4 (right) & ${\bf 13_{2}}$-${\bf 2_{5}4_{4}}$ & $6/26$ & $2^{13}=8192$\\
\hline
\end{tabular}
\caption{For each of the observables-based proofs of Figs.1-4, the fourth column shows the number of pure and hybrid bases formed by the projectors and the fifth column the total number of projectors-based parity proofs in that system.}
\label{table5} 
\end{table}

This completes our description of the procedure for generating projectors-based proofs from observables-based ones. The generation of the basis table associated with an observables-based proof (the equivalent of Table 2) takes a bit of effort, but once it is in hand the rest of the process is quite painless. A particularly simple type of observables-based proof is one in which each observable occurs in exactly two commuting sets and any two commuting sets have at most one observable in common (the proofs in Figs.1-4 are all of this type). If $O$ is the number of observables in such a proof, it is not difficult to show that the number of hybrid basis pairs is also $O$ and the number of projectors-based proofs associated with this system is $2^{O}$. Table 5 illustrates this remark by listing the number of projectors-based proofs associated with each of the observables-based proofs of Figs.1-4. The reader should be able to generate all the projectors-based proofs in these systems using the methods described in this section.

\section{\label{sec:4} KS proofs based on Kite diagrams}

\begin{figure}[htp]
\begin{center}
\includegraphics[width=0.50\textwidth]{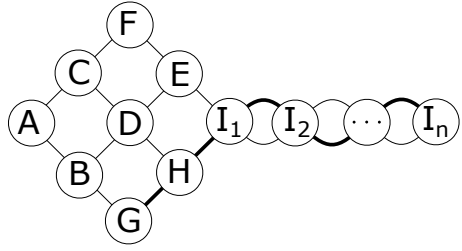}
\end{center}
\caption{The general Kite diagram. The four sets of three commuting observables, (A,C,F), (B,D,E), (A,B,G) and (C,D,H), have product {$+$\bf I} and are shown by thin lines. The other two commuting sets have $n+2$ observables each; the set (F,E,I$_{1}$,...,I$_{n}$) has product {$+$\bf I} and is shown by a thin line while the set (G,H,I$_{1}$,...,I$_{n}$) has product {$-$\bf I} and is shown by a thick line. Suitable choices of the observables A,B,...,I$_{n}$ give rise to all the proofs of the Kite family.}
\label{Fig.5}
\end{figure}

Figure 5 shows the an observables-based KS proof based on a diagram we term the Kite. There are nine observables in the body of the Kite and one of them also serves as the starting point of its tail, which can be of arbitrary length. It is easy to see that the Kite provides a KS proof because each observable occurs in two commuting sets and just one of the commuting sets has a product of {$-$\bf I}. By making suitable choices for the observables A,B,$\ldots$,H,I$_{1}$,$\ldots$,I$_{n}$, it is possible to construct KS proofs for any number of qubits from two up. Before presenting specific examples of such proofs, we draw attention to a simple class of projectors-based proofs implied by Fig.5 no matter what choices are made for the observables in it.\\

\begin{table}[ht]
\centering 
\begin{tabular}{|c | c | c | c | c |} 
\hline 
 Observables  &  +++ & + -- -- & -- + -- & -- -- +\\
\hline
A,C,F & 1 & 2 & 3 & 4 \\
B,D,E & 5 & 6 & 7 & 8 \\
A,B,G & 9 & 10 & 11 & 12 \\
C,D,H & 13 & 14 & 15 & 16 \\
\hline
\end{tabular}

\begin{tabular}{|c | c | c | c | c | c | c | c | c |} 
\hline 
 Observables  &  +++$\cdots$ & ++ -- $\cdots$ & + -- +$\cdots$ & + -- -- $\cdots$ & -- ++$\cdots$ & -- + -- $\cdots$ & -- -- +$\cdots$ & -- -- -- $\cdots$\\
\hline
F,E,I$_{1}$,$\cdots$ & 17&18&19&20&21&22&23&24 \\
G,H,I$_{1}$,$\cdots$ & 25&26&27&28&29&30&31&32  \\
\hline
\end{tabular}
\caption{Projectors defined by the Kite diagram of Fig.5. The four commuting sets of observables in the upper table define the projectors 1 through 16, while the two sets in the lower table define the groups of projectors numbered 17 through 32. The eigenvalue signatures of the projectors for the defining observables are shown at the tops of the columns. The numbers 1 through 16 represent single projectors but the numbers 17 through 32 each represent an ensemble of mutually orthogonal projectors, as explained in the text. The product of the eigenvalue signatures of each of the projectors 1 through 16 is $+$, and the same is true of each of the members of the ensembles 17 through 24. However the product of the signatures is $-$ for each of the members of the ensembles 25 through 32.}
\label{table6} 
\end{table}

To do this, we first enumerate the projectors defined by the sets of commuting observables in the Kite and then set up the basis table formed by them. These tasks are accomplished in the same manner as in Sec.2. Once the basis table is available, we point out a set of projectors-based proofs contained in it.\\

Table 6 shows the sets of commuting observables in the Kite and the projectors defined by them. The first four commuting sets define four projectors each, which are numbered from 1 to 16 and have their eigenvalue signatures indicated at the tops of the columns. The last two sets define projectors labeled by $n+2$ eigenvalue signatures each, but rather than number the projectors individually we adopt the short cut of using a single number to label all projectors having the same eigenvalues for the three observables in the body of the kite and differing only in their eigenvalues for the observables in the tail. Thus, for example, the number 17 denotes all projectors having eigenvalues $+1$,$+1$,$+1$ for F,E,I$_{1}$ but all possible combinations of eigenvalues for the observables I$_{2}$,$\cdots$,I$_{n}$ in the tail. The numbers from 18 to 32 are to be interpreted similarly.\\

\begin{table}[ht]
\centering 
\begin{tabular}{|c | c c c c c c c c | c | c c c c c c c c |} 
\hline 
Index &\multicolumn{8}{|c|}{Projectors in Basis} & Index &\multicolumn{8}{|c|}{Projectors in Basis} \\
\hline
1a &&&1&2&11&12&&&1b &&&3&4&9&10&&  \\
2a &&&1&3&15&16&&&2b &&&2&4&13&14&&  \\
3a &&&5&6&10&12&&&3b &&&7&8&9&11&&  \\
4a &&&5&7&14&16&&&4b &&&6&8&13&15&&  \\
5a &&1&4&21&22&23&24&&5b &&2&3&17&18&19&20&  \\
6a &&5&8&19&20&23&24&&6b &&6&7&17&18&21&22&  \\
7a &&9&12&29&30&31&32&&7b &&10&11&25&26&27&28&  \\
8a &&13&16&27&28&31&32&&8b &&14&15&25&26&29&30&  \\
9a &17&18&23&24&25&26&31&32 &9b& 19&20&21&22&27&28&29&30\\
\hline
\end{tabular}
\caption{Nine of the hybrid basis pairs formed by the projectors of Table 6. Complementary hybrids are shown on the same line, and distinguished by the letters a and b.}
\label{table7} 
\end{table}

\begin{table}[ht]
\centering 
\begin{tabular}{|c | c c c c c c c c c |} 
\hline 
 Index & \multicolumn{9}{|c|}{Bases in proof} \\
\hline
$1$&$1a$&$2a$&$3a$&$4a$&$5b$&$6b$&$7b$&$8b$&$9b$\\
$2$&$1a$&$2a$&$3a$&$4b$&$5b$&$6a$&$7b$&$8a$&$9a$\\
$3$&$1a$&$2a$&$3b$&$4a$&$5b$&$6a$&$7a$&$8b$&$9a$\\
$4$&$1a$&$2a$&$3b$&$4b$&$5b$&$6b$&$7a$&$8a$&$9b$\\
$5$&$1a$&$2b$&$3a$&$4a$&$5a$&$6b$&$7b$&$8a$&$9a$\\
$6$&$1a$&$2b$&$3a$&$4b$&$5a$&$6a$&$7b$&$8b$&$9b$\\
$7$&$1a$&$2b$&$3b$&$4a$&$5a$&$6a$&$7a$&$8a$&$9b$\\
$8$&$1a$&$2b$&$3b$&$4b$&$5a$&$6b$&$7a$&$8b$&$9a$\\
$9$&$1b$&$2a$&$3a$&$4a$&$5a$&$6b$&$7a$&$8b$&$9a$\\
$10$&$1b$&$2a$&$3a$&$4b$&$5a$&$6a$&$7a$&$8a$&$9b$\\
$11$&$1b$&$2a$&$3b$&$4a$&$5a$&$6a$&$7b$&$8b$&$9b$\\
$12$&$1b$&$2a$&$3b$&$4b$&$5a$&$6b$&$7b$&$8a$&$9a$\\
$13$&$1b$&$2b$&$3a$&$4a$&$5b$&$6b$&$7a$&$8a$&$9b$\\
$14$&$1b$&$2b$&$3a$&$4b$&$5b$&$6a$&$7a$&$8b$&$9a$\\
$15$&$1b$&$2b$&$3b$&$4a$&$5b$&$6a$&$7b$&$8a$&$9a$\\
$16$&$1b$&$2b$&$3b$&$4b$&$5b$&$6b$&$7b$&$8b$&$9b$\\
\hline
\end{tabular}
\caption{The 16 projectors-based KS proofs formed by the bases of Table 7 (with the bases labeled as in that table).}
\label{table8} 
\end{table}

The projectors in Table 6 form six pure bases, represented by the rows of the table. Using the orthogonality rule for projectors from different pure bases (mentioned in Sec.2), it can be verified that they form the nine pairs of complementary hybrids shown in Table 7. Table 8 shows 16 projectors-based proofs extracted from the bases of Table 7 (they can be obtained by picking one member from each of the first four hybrid pairs, which can be done in $2^{4}$ ways, and supplementing them with the required members of the remaining five hybrids). It is remarkable that any Kite diagram, irrespective of the length of its tail or the number of qubits it involves, always admits this set of 16 projectors-based proofs involving nine bases each.\footnote{Any Kite diagram actually gives rise to many more projectors-based proofs than just the 16 exhibited here. However the total number of proofs is not given by the $2^{O}$ rule because two of the commuting sets (namely, the ones that overlap along the tail) have more than one observable in common.} We believe that nine is the minimum number of bases for a projectors-based proof in any dimension, but we do not have a proof of this fact.\\

We finally give examples of $N$-qubit observables that can play the role of the labels A,B,C,$\cdots$,I$_{n}$ in Fig.5. It turns out to be sufficient to specify the members of the commuting set G,H,I$_{1}$,$\cdots$,I$_{n}$, since they determine all the other observables in the manner we explain. Table 9 lists the members of this set for three and five qubits before indicating its pattern for an arbitrary odd number of qubits, and Table 10 does the same for four, six and an arbitrary even number of qubits. In each case the commuting observables are displayed horizontally, with their corresponding qubits vertically aligned. The observables G and H are always the ones whose first qubits are in bold, while the others are I$_{1}$,$\cdots$,I$_{n}$ (it is immaterial how the associations are made within these two groups). Once these observables have been placed at their locations in the Kite, F and E are chosen as the observables obtained from G and H by swapping their first qubits (which are always X and Z). The observables A,B,C and D are then uniquely determined by the requirement that they have the necessary commutation and product properties. Fig.6 (left) shows the Kite proof that results on applying this procedure to the 3-qubit observables of Table 9. The 4-qubit observables of Table 10 yield a Kite diagram with a tail of length two, but rather than exhibiting this proof we show, in Fig.6 (right), a more economical proof with a tail of length one.

\begin{table}[ht]
\centering
\begin{tabular}{ccc}
$\textbf{Z}$ & $I$ & $Z$ \\
$I$ & $Z$ & $Z$ \\
$\textbf{X}$ & $X$ & $X$ \\
$Y$ & $Y$ & $X$ \\
\end{tabular} \label{KiteOdd3}
\qquad
\begin{tabular}{ccccc}
$\textbf{Z}$ & $I$ & $I$ & $I$ & $Z$ \\
$I$ & $Z$ & $I$ & $I$ & $Z$ \\
$I$ & $I$ & $Z$ & $I$ & $Z$ \\
$I$ & $I$ & $I$ & $Z$ & $Z$ \\
$\textbf{X}$ & $X$ & $X$ & $X$ & $X$ \\
$Y$ & $Y$ & $Y$ & $Y$ & $X$ \\
\end{tabular} \label{KiteOdd5}
\qquad
\begin{tabular}{ccccc}
$\textbf{Z}$ & $I$ & $\cdots$ & $I$ & $Z$ \\
$I$ & $Z$ & $\ddots$ & $\vdots$ & $\vdots$ \\
$\vdots$ & $\ddots$ & $\ddots$  & $I$ & $Z$ \\
$I$ & $\cdots$  & $I$ & $Z$ & $Z$ \\
$\textbf{X}$ & $\cdots$ & $X$ & $X$ & $X$ \\
$Y$ & $\cdots$ & $Y$ & $Y$ & $X$ \\
\end{tabular} \label{KiteOddN}
\qquad
\caption{Observables for a Kite proof based on three qubits (left), five qubits (middle) and an arbitrary odd number of qubits (right). }\label{KiteOdd}
\end{table}

\begin{table}[ht]
\centering
\begin{tabular}{cccc}
$\textbf{Z}$ & $Z$ & $Z$ & $Z$ \\
$Y$ & $Y$ & $Z$ & $Z$ \\
$\textbf{X}$ & $I$ & $X$ & $I$ \\
$I$ & $X$ & $I$ & $X$ \\
$I$ & $I$ & $X$ & $X$ \\
\end{tabular} \label{KiteEven4}
\qquad
\begin{tabular}{cccccc}
$\textbf{Z}$ & $Z$ & $Z$ & $Z$ & $Z$ & $Z$ \\
$Y$ & $Y$ & $Z$ & $Z$ & $Z$ & $Z$ \\
$\textbf{X}$ & $I$ & $X$ & $I$ & $I$ & $I$ \\
$I$ & $X$ & $I$ & $X$ & $I$ & $I$ \\
$I$ & $I$ & $X$ & $I$ & $X$ & $I$ \\
$I$ & $I$ & $I$ & $X$ & $I$ & $X$ \\
$I$ & $I$ & $I$ & $I$ & $X$ & $X$ \\
\end{tabular} \label{KiteEven6}
\qquad
\begin{tabular}{cccccc}
 & & & & & \\
$\textbf{Z}$ & $Z$ & $Z$ & $Z$ & $\cdots$ & $Z$ \\
$Y$ & $Y$ & $Z$ & $Z$ & $\cdots$ & $Z$ \\
$\textbf{X}$ & $I$ & $X$ & $I$ & $\cdots$ & $I$ \\
$I$ & $X$ & $I$ & $X$ & $\ddots$ & $\vdots$ \\
$\vdots$ & $\ddots$ & $\ddots$ & $\ddots$ & $\ddots$ & $I$ \\
$I$ & $\cdots$ & $I$ & $X$ & $I$ & $X$ \\
$I$ & $\cdots$ & $I$ & $I$ & $X$ & $X$ \\
\end{tabular} \label{KiteEvenN}
\qquad
\caption{Observables for a Kite proof based on four qubits (top left), six qubits (top right) and an arbitrary even number of qubits (bottom)}\label{KiteEven}
\end{table}

\begin{figure}
\centering
\begin{tabular}{cc}
\epsfig{file=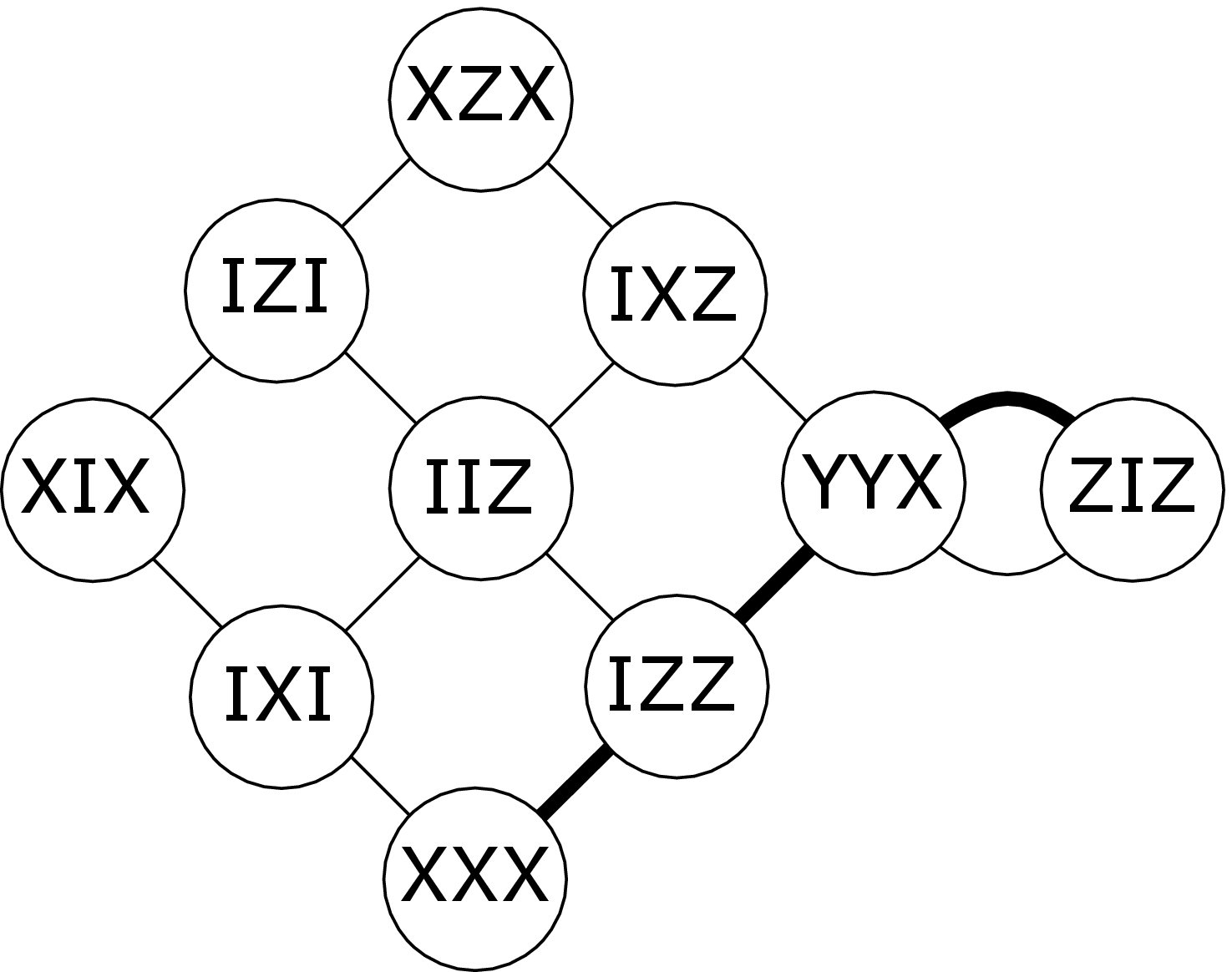,width=0.4\linewidth,clip=} &
\epsfig{file=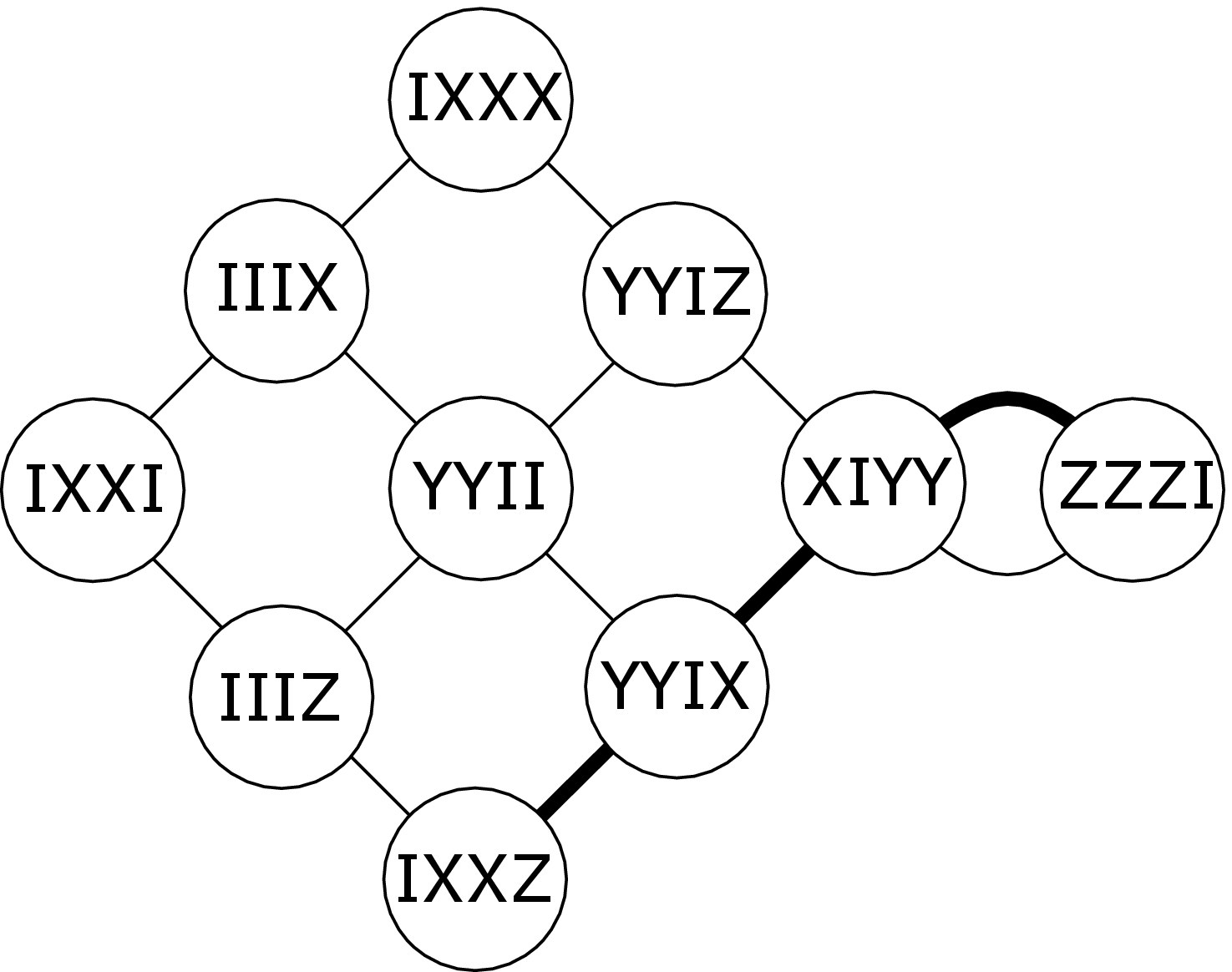,width=0.4\linewidth,clip=} \\
\end{tabular}
\caption{3-qubit Kite (left) based on the observables of Table 9 and a 4-qubit Kite (right) based on the same skeleton as at the left.}
\label{Fig.6}
\end{figure}

The economy we pointed out for the 4-qubit proof extends to higher qubit proofs as well: we have found $N$-qubit proofs, for all $N \geq 3$, whose longest commuting sets involve considerably less than the $N+1$ observables involved in the proofs of Tables 9 and 10. Some examples of such  proofs are shown in Table 11. The 16-qubit proof has a commuting set of just 7 observables, which is much less than the 17 involved in the proof of Table 10. This sort of compression allows for the design of much more economical KS tests as one goes to a large number of qubits.

\begin{table}[ht]
\centering
\begin{tabular}{ccccccc}
$\textbf{Z}$ & $Z$ & $Z$ & $Z$ & $Z$ & $I$ & $I$ \\
$\textbf{X}$ & $X$ & $Z$ & $X$ & $X$ & $Z$ & $Z$ \\
$Y$ & $I$ & $X$ & $Z$ & $Z$ & $X$ & $X$ \\
$I$ & $Y$ & $I$ & $X$ & $I$ & $Z$ & $X $ \\
$I$ & $I$ & $X$ & $I$ & $X$ & $X$ & $Z $ \\
\end{tabular} \label{M5N7}
\qquad
\begin{tabular}{ccccccccccc}
&&&&&&&&&&\\
$ I$ & $I$ & $Z$ & $Z$ & $Z$ & $Z$ & $Z$ & $Z$ & $Z$ & $Z$ & $Z  $ \\
$ \textbf{Z}$ & $I$ & $Z$ & $Z$ & $Z$ & $X$ & $X$ & $I$ & $X$ & $X$ & $I $ \\
$ I$ & $Z$ & $X$ & $X$ & $I$ & $Z$ & $Z$ & $Z$ & $I$ & $I$ & $I $ \\
$ \textbf{X}$ & $I$ & $X$ & $I$ & $X$ & $X$ & $I$ & $X$ & $Z$ & $X$ & $X $ \\
$ I$ & $X$ & $I$ & $X$ & $I$ & $I$ & $X$ & $I$ & $I$ & $Z$ & $Z $ \\
$ Y$ & $Y$ & $I$ & $I$ & $X$ & $I$ & $I$ & $X$ & $X$ & $I$ & $X $ \\
\end{tabular} \label{M6N11}
\qquad
\begin{tabular}{cccccccccccccccc}
&&&&&&&&&&\\
$\textbf{Z}$ & $Z$ & $Z$ & $Z$ & $Z$ & $Z$ & $Z$ & $Z$ & $I$ & $I$ & $I$ & $I$ & $I$ & $I$ & $I$ & $I $ \\
$\textbf{X}$ & $X$ & $X$ & $X$ & $I$ & $I$ & $I$ & $I$ & $Z$ & $Z$ & $Z$ & $Z$ & $I$ & $I$ & $I$ & $I $ \\
$Y$ & $I$ & $I$ & $I$ & $X$ & $X$ & $X$ & $I$ & $X$ & $I$ & $I$ & $I$ & $Z$ & $Z$ & $I$ & $I $ \\
$I$ & $Y$ & $I$ & $I$ & $Y$ & $I$ & $I$ & $I$ & $I$ & $X$ & $X$ & $X$ & $X$ & $I$ & $Z$ & $I $ \\
$I$ & $I$ & $I$ & $I$ & $I$ & $Y$ & $I$ & $X$ & $Y$ & $Y$ & $I$ & $I$ & $I$ & $I$ & $X$ & $Z $ \\
$I$ & $I$ & $Y$ & $I$ & $I$ & $I$ & $I$ & $Y$ & $I$ & $I$ & $Y$ & $I$ & $Y$ & $X$ & $I$ & $X $ \\
$I$ & $I$ & $I$ & $Y$ & $I$ & $I$ & $Y$ & $I$ & $I$ & $I$ & $I$ & $Y$ & $I$ & $Y$ & $Y$ & $Y $ \\
\end{tabular}\label{Kite7_16}
\qquad
\caption{Kite proofs for 7 qubits based on 5 commuting observables (top), for 11 qubits based on 6 commuting observables (middle) and for 16 qubits based on 7 commuting observables (bottom). The full proofs can be constructed by placing these observables on the skeleton of the Kite in the manner explained in the text.}\label{ShortKites}
\end{table}

\section{\label{sec:5} Discussion}

This paper has established the following results:\\

\noindent
1. The hierarchy of $N$-qubit Pauli groups (for $N \geq 2$) contains many subsets of observables that provide parity proofs of the KS theorem. We restrict our attention to proofs that are irreducible (i.e. that cannot be reduced to simpler proofs by omitting some subset of observables and/or qubits in them) and unitarily inequivalent. With these caveats, the 2-qubit group gives rise to only two distinct types of proofs (the Peres-Mermin square and a more complicated structure we call the ``whorl", which are shown as Figs.1 and 2 of Ref. \cite{Waegell2012a}), but the 3-qubit group leads to many more \cite{Waegell2012a} and the possibilities increase rapidly as one goes upwards in the number of qubits. It would be an interesting problem to make a more systematic inventory of the proofs at a given value of $N$ and to get a feeling for how this number grows with $N$. There are two devices we have introduced in this paper that could assist with this task: the first is a diagrammatic representation of each proof (from which it can be verified by inspection) and the second is a symbol that captures the important features of the observables and special commuting sets in the proof. However we should caution that neither of these devices suffices to pin down a proof uniquely since there are distinct proofs that can be accommodated on the same diagrammatic skeleton, as well as inequivalent proofs that share the same symbol (see Fig.7 of Ref. \cite{Waegell2012a} for an example). Even in the absence of a detailed knowledge of the terrain, one can state quite confidently that the Pauli group (particularly as one goes to a larger number of qubits) abounds in a great variety of structures that can be used to give transparent demonstrations of quantum contextuality.\\

\noindent
2. The second major point of this paper is that every observables-based KS proof can be used to generate a system of projectors and bases from which a a large number of projectors-based KS proofs can be obtained. The algorithm for generating these proofs is simple, and the proofs themselves are easy to validate because only a simple parity check is called for. We have introduced a detailed symbol for a projectors-based proof that describes the projectors and bases in it, both as a way of summarizing the key aspects of the proof and to draw attention to the wide variety of proofs that can coexist within the same framework of pure and hybrid bases provided by an observables-based proof. Since the observables-based proofs are themselves very numerous, and each spawns a large number of projectors-based proofs (typically thousands), the quantity and variety of the latter proofs vastly outstrip those of the former. We should add that the symbols we have introduced for the projectors-based proofs, though useful and informative, do not serve to pin them down uniquely since we have found many examples of inequivalent proofs that are described by the same symbol.
Among all the observables-based and related projectors-based proofs, there is a simple class that is worth singling out for special mention: it is the one in which each observable occurs in exactly two commuting sets and any two commuting sets have at most one observable in common. One knows in this case, even before one has set up the basis table, that one will find exactly $2^{O}$ projectors-based proofs.\\

\noindent
3. We have discovered several infinite families of observables-based proofs (that we term the Star, the Wheel, the Whorl and the Kite) whose members yield KS proofs for all numbers of qubits from two up. The Kite is the most complex of these families, and we have given a detailed discussion of it in this paper. This family has two remarkable features. The first is that for a system of $N$ qubits it is always possible to find commuting sets of considerably less than $N$ observables that can be used to construct a KS proof, thus leading to greater economy in the design of experimental tests based on this class of proofs. And the second is that \emph{any} Kite proof always yields a projectors-based proof involving only nine bases (or experimental contexts). These proofs are generalizations of the classic 18-9 proof of Cabello et al \cite{Cabello1996} in four dimensions and hold in all dimensions of the form $2^{N}$, for $N \geq 2$. It is an open question whether there are any proofs involving less than nine bases in these dimensions (we believe the answer is no). An even more basic question is whether there are any projectors-based parity proofs\footnote{Non parity proofs based on rays are known in all dimensions $\geq 3$ \cite{CEG2005,ZP}.} at all in any even dimension not of the form $2^{N}$. Again we suspect that the answer is no, but we do not have a proof of this conjecture (and would find it fascinating if someone came up with a counterexample).\\

There is currently a great interest in contextuality and nonlocality and their experimental tests. In one interesting development, it has been shown that any projectors-based KS proof can be converted into an inequality for testing quantum contextuality \cite{Cabello2008} and a number of experimental tests of such inequalities have actually been carried out \cite{Kirchmair}. A number of works have shed light on contextuality in a manner that bypasses the KS theorem \cite{Klyachko}. On the formal side, a connection between KS proofs and ``logical Bell inequalities" has been made in Ref. \cite{Abram} and the question of state-independent contextuality for identical particles has been explored in Ref. \cite{Cabello2013}. Although the present work concentrates entirely on qubits, it should be pointed out that KS proofs for qudits (i.e., $d$-state systems) have been explored in Ref. \cite{Cerf}. Among the practical applications of contextuality that are receiving attention are quantum key distribution \cite{BPPeres}, quantum error correction  \cite{Hu2008,Raussendorf2001}, random number generation \cite{Svozil}, parity oblivious transfer \cite{Spekkens} and the design of relational databases \cite{Abramsky2012}. This is not a complete survey, of course, but should convey a feeling for the broader context in which the results reported here may be of interest.


\end{document}